\documentstyle[a4,11pt,epsfig,epsf]{article}

\columnwidth\textwidth

\def\n.c.#1#2#3{         {\it Nuovo Cim. }{\bf #1} (19#2) #3}
\def\r.n.c.#1#2#3{       {\it Riv. del Nuovo Cim. }{\bf #1} (19#2) #3}

\def\be{\begin{equation}}       
\def\ee{\end{equation}}
\def\bear{\be\begin{array}}
\def\eear{\end{array}\ee}
\def\bea{\begin{eqnarray}}
\def\eea{\end{eqnarray}}

\def\21{$SU(2) \ot U(1)$}
\def\ot{\otimes}

\def\bold#1{\setbox0=\hbox{$#1$}
     \kern-.025em\copy0\kern-\wd0
     \kern.05em\copy0\kern-\wd0
     \kern-.025em\raise.0433em\box0 }

\newcommand{\lsim}{\mbox{\raisebox{-1.ex}
{$\stackrel{\textstyle <}{\textstyle \sim}$}}}
\newcommand{\gsim}{\mbox{\raisebox{-1.ex}
{$\stackrel{\textstyle >}{\textstyle \sim}$}}}
\newcommand{\sstt}      {\sin^2 2\theta}
\newcommand{\sinsq}      {\sstt}
\newcommand{\dms}       {\Delta m^2}
\newcommand{\dmsq}       {\dms}
\newcommand{\nue}       {{\nu}_{\rm e}}
\newcommand{\nuebar}    {\bar{\nu}_{\rm e}}
\newcommand{\numu}      {\nu_{\rm \mu}}

\newcommand{\nutau}     {{\nu}_{\rm \tau}}

\newcommand{\units}[1] {\mbox{$\textstyle #1$}}

\begin{document}
\begin{flushright}
FTUV/IFIC-9910\\
hep-ph/9902339
\end{flushright}

\begin{center}
\phantom{pepe}
\vspace{2cm}

\Large{\bf A Review on  Neutrino Physics, Mass and Oscillations.
}
\footnote{Revised and expanded version of a Talk at the ``Instituto de 
Fisica Daza de Valdes'', CSIC , Madrid, November 1998} 
\end{center}

\begin{center}
 E. Torrente-Lujan. \\
{\it  IFIC- CSIC,
Dept. de Fisica Teorica\\ Universidad de Valencia,
46100 Burjassot, Valencia, Spain.\\
e-mail: e.torrente@cern.ch.}
\end{center}

\begin{abstract}
We present  a short review of the present status of the problem of 
neutrino masses and mixing including  a survey of 
theoretical motivations and models, experimental
 searches and implications of recently appeared  solar and 
atmospheric neutrino data, which strongly indicate 
nonzero neutrino masses. Such data apparently requires the 
existence of light sterile neutrino in addition to the 
three active ones and two-generation nearly maximal mixing.
\end{abstract}

\newpage

\section{Introduction}
Among all the fundamental particles the neutrino occupies a
unique place in many ways and as such it has shed light on many 
important aspects of our present understanding of nature and is 
believed to hold the key to the physics beyond the 
Standard Model.
Presently, the question of whether the neutrino has mass is one of the 
outstanding issues in  particle physics, astrophysics, 
cosmology and theoretical physics in general.
There are several theoretical, observational and experimental motivations which
justify the searching for possible non-zero neutrino 
masses (see i.e. \cite{langacker1,fuk2} for excellent older 
reviews on this matter).

Fermion masses in general are one of the major problems of the Standard
 Model. Observation or non-observation of neutrino masses could introduce 
an useful new perspective on the subject.
If massless they would be the only fermions with this property. 
A property which is not dictated by a fundamental underlying 
principle (at least one which is known presently), such as gauge invariance in the  case of the photon.
If massive the question is why are their masses so much smaller than those of their charged partners.
Although theory alone can not predict neutrino masses, it is certainly true 
that they are strongly suggested by present theoretical models of
elementary particles and 
most extensions of the Standard Model definitively require 
neutrinos to be massive.
They therefore constitute a powerful probe of new physics at a scale 
larger than the electroweak scale.

Some  hints
at accelerator experiments as the  
the observed indications of spectral distortion and deficit of solar 
neutrinos and the ratio of atmospheric $\nu_e/\nu_\mu$ neutrinos and 
their zenith distribution 
are naturally accounted by the oscillations of a massive neutrino.
Recent claims of the High-statistics high precision SuperKamiokande 
experiment are unambiguous and left little room for the scepticism.

Neutrinos are basic ingredients of astrophysics and cosmology.
There may be a hot dark matter component (HDM) to the Universe:
simulations of structure
formation fit the observations only when one has some 
20 \% of HDM, 
the best fit being two neutrinos with a total mass of 4.7 eV. 
If so, neutrinos would be, at least in quantity, one of the most important 
ingredients in the Universe.
There seems to be however 
some kind of conflict within cosmology itself: observations of
distant objects favor a large cosmological constant instead of HDM.

Regardless of  mass and oscillations, 
astrophysical interest in the neutrino and their properties 
arises from the fact that it is copiously 
produced  in high temperature and/or high density environment and 
it often dominates the physics of those astrophysical objects. The 
interactions of the neutrino with matter is so weak that it passes
 freely through any ordinary matter existing in the Universe. This makes 
neutrinos to be a very efficient carrier of energy drain from optically 
thick objects. At the same time they give a good probe for the interior of 
such dense objects.
For example, the solar neutrino flux is, together with heliosysmology, 
 one of the two known probes of the solar core. A similar statement 
 applies to the type-II supernovas: The most interesting questions 
around supernovas, the explosion dynamics itself with the 
shock revival, and,  the synthesis of the heaviest elements by 
 r-process, could be positively affected by changes in the neutrino flux, 
e.g. by MSW active or sterile conversions. Finally, ultra 
high energy neutrinos are called to be useful probes of diverse distant 
astrophysical objects. Active Galactic Nuclei (AGN) should be copious 
 emitters of $\nu$'s, providing both detectable point sources and an 
 observable diffuse background which is larger in fact than the atmospheric 
 neutrino background in the very high energy range.

\section{The neutrinos in the Standard Model.}

The standard model (SM)  contains three left-handed neutrinos. The three neutrinos are 
represented by two-component Weyl spinors
$\nu_i, i=e,\mu,\tau$ each describing a left-handed 
 fermion. They are upper components of weak isodoublets $L_i$, they 
have $I_{3W}=1/2$ and an unit of the global ith lepton number.
These standard model neutrinos are strictly massless. 
The only 
Lorenz scalar made out of them is the Majorana mass, of the form 
$\nu^t_i \nu_i$; it has the quantum number of a weak isotriplet, with 
$I_{3W}=1$ as well as two units of total lepton number. Thus to generate 
a renormalizable 
Majorana mass term at the tree level one needs a Higgs isotriplet 
 with two units of lepton number. Since in the strict standard model 
 Higgs is a weak isodoublet Higgs, there are no tree-level neutrino masses.
If we want to consider quantum corrections we should consider 
 effective terms where a weak isotriplet is made out of two isodoublets and 
 which are not invariant under lepton number symmetry. The conclusion is that the 
standard model neutrinos are kept massless by global chiral lepton number 
 symmetry
(and more general properties as renormalizability of the theory).
However this is a rather formal conclusion, there is no any other independent, compelling theoretical 
argument in favor of such symmetry. Or, with other words,
 there is not any reason why we would like to 
keep it intact.

Independent from mass oddities in any other respect
 neutrinos are very well behaved particles within the SM framework.
Some properties are unambiguously known about neutrinos. 
The LEP  line-shape measurement imply that are 
only three ordinary light neutrinos.
Big Bang Nucleosynthesis (BBN) constrains the parameters of 
possible sterile neutrinos (which interact and are produced only by mixing). 
{\em All the existing} data on the  the weak interaction
processes in which neutrinos take part are perfectly described by the 
Standard Model 
charged-current (CC) and neutral-current (NC) Lagrangians:
\begin{eqnarray}
L_I^{CC}&=&-\frac{g}{\surd 2} \sum_{l=e,\mu,\tau} 
\overline{\nu_{lL}}\gamma_\alpha l_L W^\alpha+ h.c.\\
L_I^{NC}&=&-\frac{g}{2 \cos \theta_W} \sum_{l=e,\mu,\tau} 
\overline{\nu_{lL}}\gamma_\alpha \nu_{lL} Z^\alpha + h.c.
\end{eqnarray}
The CC and NC interaction Lagrangians  conserve the 
total three lepton numbers $L_{e,\mu,\tau}$ while CC interactions determine 
the notion of flavor neutrinos $\nu_{e,\mu,\tau}$. 
There are no indications in favor of violation of lepton numbers in weak
 processes. From the existing experiments very strong bounds 
on branching ratios of rare, lepton number violating,
 processes are obtained, for
example: 
$R(\mu\to e\gamma)< 5\times 10^{-11}$ and $ R(\mu\to 3 e)< 10^{-12}$
(\cite{PDG98}, 90\% CL).

\section{Neutrino mass terms and models.}

Any satisfactory 
model that generates neutrino masses must contain a natural 
mechanism that explains their small value, relative to that of their
charged partners.
Given the latest experimental indications it would also  be desirable  that
includes justification for light sterile neutrinos and near maximal 
mixing.
To generate neutrino masses without new fermions, we must break 
lepton number by  adding to the 
SM Higgs Fields  carrying lepton numbers, one can 
arrange then to break lepton number 
explicitly or spontaneously through their interactions.
Possibly, the most familiar approach
 to give neutrino masses is, however, to introduce for 
each one an electroweak neutral singlet. 
This happens naturally in left-right symmetric models where the origin of 
SM parity violation is ascribed to the Spontaneous breaking of a B-L
symmetry.
In the SO(10) GUT the Majorana neutral particle  N
enters in a natural way in order to complete the 
matter multiplet, 
the neutral N is a $SU(3)\times SU(2)\times U(1) $ singlet.
According to the scale where they have relevant effects, Unification 
(i.e. the aforementioned SO(10) GUT) and 
weak-scale approaches (i.e. radiative models) can be distinguished.

Phenomenologically, mass terms can be viewed as describing 
 transitions between right (R) and left (L)-handed
states. 
For a set of four fields: $\psi_L,\psi_R,(\psi^c)_L,(\psi^c)_R$, the 
most general mass part of the free-field Lagrangian can be 
written as:
\begin{eqnarray}
-L&=& 
m_D \left ( \overline{\psi}_L \psi_R\right ) 
+\frac{1}{2} m_T \left ( \overline{(\psi_L)^c} \psi_L\right ) 
+\frac{1}{2} m_S \left ( \overline{(\psi_R)^c} \psi_R\right )+h.c. 
\label{e2001}
\end{eqnarray} 
In terms of the  new Majorana fields:
$\nu=(1/\surd 2) (\psi_L+(\psi_L)^c)$, 
$N=(1/\surd 2) (\psi_R+(\psi_R)^c)$, the Lagrangian becomes:
\begin{eqnarray}
-L&=& \left ( \overline{\nu}, \overline{N}\right ) 
\pmatrix{m_T & m_D \cr m_D & m_S}
\pmatrix{\nu \cr N}
\label{e2003}
\end{eqnarray}
where the neutrino mass matrix is evident. 
Diagonalizing this matrix one finds that the physical 
particle content is given by two Majorana mass
eigenstates: The inclusion of the Majorana mass splits the four 
degenerate states of the Dirac field into two non-degenerate Majorana
 pairs.
If we assume that the states $\nu,N$ 
are respectively active and sterile, the ''Majorana masses'' $m_T$ and 
$m_S$ transform as weak triplets and singlets respectively.
$m_D$ is an standard Dirac mass term. 
The neutrino  mass matrix can easily be generalized to three 
or more families, in which
case the masses become matrices themselves.
The complete flavor mixing comes from two different 
parts, the diagonalization of 
 the charged lepton Yukawa couplings and that of the neutrino masses. 
In most simple extensions of the standard model, this CKM-like leptonic 
 mixing  is totally 
arbitrary with parameters only to be determined by experiment. Their 
 prediction, as for for the quark hierarchies and mixing, needs 
 further theoretical assumptions (i.e. Ref.\cite{RAMOND} predicting
$\nu_\mu-\nu_\tau$ maximal mixing).

The case $m_T,m_S\equiv 0$ in Eq.(\ref{e2003}) 
corresponds to a purely Dirac mass term. In this case
$\nu,N$ are degenerate with mass $m_D$ and a  four component 
Dirac field can be  recovered as $\nu\equiv \nu+N$.
The Dirac mass term allows a conserved lepton number $L=L_\nu+L_N$.
For an ordinary Dirac neutrino the $\nu_L$ is active 
and $\nu_R$ is weak sterile, an SU(2) singlet, 
the mass term describes then a $\Delta I=1/2$ transition and 
 is generated from  SU(2) breaking with a Yukawa coupling:
\begin{eqnarray}
-L_{Yuk}&=&h_\nu \left (\overline{\nu_l},\overline{l}\right )_L 
\pmatrix{\phi^0\cr \phi^-} N_R + h.c.
\label{e2005}
\end{eqnarray}
One has $m_D=h_\nu v/2$ where v is the vacuum expectation value of the 
Higgs doublet and $h_\nu$ is the corresponding Yukawa coupling.
A neutrino Dirac mass is qualitatively just like any other fermion 
masses, but that leads 
to the question of why it is so small in comparison with the 
rest of fermion masses: one would require
 $h_{\nu_e}<10^{-10}$ in order to have $m_{\nu e}< 10 $ eV.
Or in other words: $h_{nu_e}/h_e\sim 10^{-5}$ while
 for the hadronic sector we have 
 $h_{up}/h_{down}\sim O(1)$.


A pure Majorana mass transition term, 
$m_T$ or $m_S$ terms in  Lagrangian (\ref{e2003}), describes in fact
 a particle-antiparticle transition
 violating lepton number by two units 
($\Delta L=\pm 2$). It can be viewed as the creation or annihilation of 
two neutrinos leading therefore to neutrinoless double beta decay.

For N  a gauge singlet, 
a renormalizable mass term of the type $L_N=m_S N^t N$ is
allowed by SM SU(3)xSU(2)xU(1) symmetry.
However, it would not be consistent in general with 
unified symmetries, i.e. with a full SO(10) symmetry.
In such theory the most straightforward possibility for generating a 
term like it in the full theory would be to include a 
{\bf 126} Higgs and a Yukawa coupling. Alternatively, one can imagine
 more complicated interactions containing products of several simpler Higgs
 fields.
Whatever the concrete model, the main point to have into account 
is that, it is strongly suggested that $m_S$ is associated 
 with breaking of some unified symmetry, the expected scale for it should be
 in a large  range covering from $\sim$ TeV (left-right models) to 
GUT scales $\sim 10^{15}-10^{17}$ GeV (from couplings unification).

If $\nu_L$ is active then $\Delta I$=1 and $m_T$ must be generated by either 
an elementary Higgs triplet or by an effective operator involving two 
Higgs doublets arranged to transform as a triplet. 

For an elementary triplet $m_T\sim h_T v_T$, where $h_T$ is a Yukawa
 coupling and $v_T$ is the triplet VEV.
The simplest implementation (the old  Gelmini-Roncadelli model)
 is excluded by the LEP data on the Z width: the corresponding 
Majoron couples to the Z boson increasing significantly its width. 
Variant models involving 
explicit lepton number violation or in which the Majoron is mainly a weak 
 singlet (invisible Majoron models) could still be possible.

For an effective operator originated mass, one expects 
$m_T\sim 1/M$ where  M is the scale 
of the new physics which generates the operator. 
One can see this easily in the
 see-saw scheme:
The N's communicate with the familiar fermions through the 
Yukawa interactions (Eq.(\ref{e2005}).
If N were  massless the effect of Eq.(\ref{e2005}) 
would be  to generate masses 
of the same order as ordinary quark and lepton masses.
If N are massive enough Yukawa and mass terms 
 could be integrated  out to generate an  
effective term of the type (L, lepton doublet):
$$ L_{eff}\sim h^2/m_S L L \phi \phi +h.c.$$ 

In the seesaw limit in Eq.(\ref{e2003}), taking $m_T\sim 1/m_S \sim 0, m_D<< m_S\sim GUT$,   the
two Majorana neutrinos acquire respectively masses
$m_1  \sim  m_D^2/m_S<< m_D$ ,$ m_2  \sim  m_S$.
There is one heavy neutrino and one neutrino much lighter than the typical
 Dirac fermion mass. This is a natural way of generating two well 
 separated  mass scales.

Now we come again to the Majorana mass $m_S$ for the sterile neutral N:
$m_S$ can vary anywhere from the TeV scale to the Planck Scale among the 
large quantity of proposed concrete seesaw and related models.
The TeV scale models are motivated, i.e., by left-right symmetric 
models. With typical $m_D$'s, one expects masses of order 
 $10^{-1}$ eV, 10 keV, and 1 MeV for the 
$\nu_{e,\mu,\tau}$ respectively violating cosmological bounds 
unless heavy neutrinos decay rapidly and invisibly.
GUT motivated
intermediate scales ($10^{12}-10^{16}$ GeV) yield 
typical masses in the range relevant to hot dark matter, and solar and 
atmospheric neutrino oscillations.
For $m_S\sim 10^{12}$ GeV 
(some superstring models, GUT with 
multiple breaking stages) one can obtain light neutrino masses of the order
$(10^{-7}$ eV, $10^{-3}$ eV, 10 eV). Such range of masses  would 
allow the interpretation of the Solar and 
atmospheric deficits as, respectively, $\nu_e\to \nu_\mu$ , $\nu_\mu\to \nu_\tau$ oscillations. $\nu_\tau$ could be considered as a dark matter candidate.
For $m_S\sim 10^{16}$ ( grand unified seesaw with large Higgs representations)
one typically finds smaller masses around 
$(10^{-11}$, $10^{-7}$, $10^{-2}$) eV 
somehow more difficult to fit into the present known experimental facts.

Models have been proposed where small tree level neutrino masses are obtained
 without making use of large scales.
The model proposed by \cite{valle1} (inspired  by previous 
superstring models)
offers an example of the possibility of 
having neutrino mass matrices more general 
than that given by Eq.(\ref{e2003}).
The incorporation of 
additional iso-singlet neutral fermions $N_i$  leads
 to a matrix of the type:
\begin{equation}
\pmatrix{0 & m_D & 0\cr
m_D & 0 & M \cr
0 & M& \mu }.
\label{e3004}
\end{equation}
The
smallness of neutrino masses is explained directly from the, otherwise
 left unexplained, smallness of the 
parameter  $\mu$ in such a model. Moreover, there would be  neutrino mixing even if the light 
neutrino remains strictly massless ($\mu\equiv 0$).

The anomalies observed in the solar neutrino flux, atmospheric flux and 
low energy accelerator experiments cannot all be explained 
consistently without introducing a light, then neccesarily sterile, 
 neutrino.
If all the Majorana masses are small, active neutrinos 
can oscillate into the sterile right handed fields. 
Light sterile neutrinos can appear in particular see-saw mechanisms 
if additional assumptions are considered 
(``singular see-saw `` models) with some unavoidable fine tuning.
The alternative to such fine tuning would be seesaw-like suppression 
for sterile neutrinos  involving  new unknown
 interactions, i.e. family symmetries,
 resulting in substantial additions to the SM,
(i.e. some sophisticated superstring-inspired models, Ref.\cite{benakli}).

Finally, weak scale, radiative generated mass  
models where the neutrino masses are zero at tree level 
constitute a very different class of models: they explain
in principle the smallness of $m_\nu$ for both active and sterile 
neutrinos. Different mass scales are generated naturally by different 
number of loops involved in generating each of them. 
The actual implementation
generally  requires however the ad-hoc introduction of new Higgs 
 particles  with nonstandard electroweak quantum numbers and lepton
 number-violating couplings \cite{valle2}.

The magnetic dipole moment  is another 
probe of possible new interactions.
Majorana neutrinos 
have identically zero magnetic and electric dipole moments.
Flavor transition magnetic moments are  allowed however in general for both 
Dirac and Majorana neutrinos. 
Limits obtained 
 from laboratory experiments are of the order of a few  
$\times 10^{-10}\mu_B$ and those from stellar physics or cosmology are 
$O(10^{-11}-10^{-13})\mu_B$. 
 In the SM electroweak theory, extended to allow for Dirac neutrino masses, 
the neutrino magnetic dipole moment is nonzero and given, as 
(\cite{PDG98} and references therein):
\begin{eqnarray}
\mu_\nu&=& \frac{3 e G_F m_\nu}{8 \pi^2 \surd 2}= 3\times 10^{-19} (m_\nu/1\ eV)\mu_B
\end{eqnarray}
where $\mu_B$ is the Bohr magneton. 
The proportionality of $\mu_\nu$ to the 
neutrino mass is due to the absence of any interaction of $\nu_R$ other 
than its Yukawa coupling which generates its mass. In left-right 
symmetric theories $\mu_\nu$ is proportional to the charged 
lepton mass: a value  of $\mu_\nu\sim 10^{-13}-10^{-14}\mu_B$ can be reached 
still too small to have practical astrophysical consequences.

Magnetic moment interactions arise in any renormalizable gauge theory
only as finite  radiative corrections. The diagrams which generate a magnetic 
moment will also contribute to the neutrino mass once the external 
photon line is removed. 
In the absence of additional symmetries a large magnetic moment is 
incompatible with a small neutrino mass.
The way out suggested by Voloshin  consists in defining a 
SU(2)$_\nu$ symmetry acting on the space $(\nu,\nu^c)$, magnetic moment 
terms are singlets under this symmetry. In the limit of 
exact SU(2)$_\nu$ the neutrino mass is forbidden but $\mu_\nu$ is 
allowed. Diverse concrete models have been 
proposed where such symmetry  is embedded into an 
extension of the SM (left-right symmetries,  SUSY with horizontal gauge 
symmetries \cite{babu1}).

\section{Experimental considerations.}
\subsection{Laboratory, reactor and accelerator results.}

No indications in favor of non-zero neutrino masses were found in direct
kinematical searches:
\begin{enumerate}
\item From the measurement of the high energy part of the 
$\beta$ spectrum in the tritium decay: 
The Troitsk  and Mainz experiments obtain respectively $m_{\nu e}< 3.4$ eV \cite{troitsk} and 
Mainz  $m_{\nu e}< 2.7$ eV \cite{mainz}. Both measurements are plagued by 
interpretation ambiguities: apparition of negative 
 mass squared and bumps at the end of the spectrum. 

\item Limits for the muon neutrino mass have been derived using the decay 
channel $\pi^+\to\mu^+ \nu_\mu$ at intermediate energy accelerators (PSI, LANL).
The present limits are $m_{\nu \mu}\lsim 160 $ keV \cite{PSI}.

\item 
A tau neutrino mass of less than 30 MeV is well established and confirmed by several experiments:
limits of 28, 30 and 31 MeV have also been  obtained by the OPAL, CLEO
 and ARGUS experiments respectively \cite{OPAL}.  
The best upper limit for the $\tau$ neutrino mass has been derived using the 
decay mode $\tau\to 5 \pi^\pm \nu_\tau$ by the ALEPH collaboration 
\cite{ALEPH}: 
$m_{\nu\tau}<$ 18 MeV (95\% CL).
\end{enumerate}

Many experiments on the search for neutrinoless double-beta decay
[$(\beta\beta)_{0\nu}$], 
$$(A,Z)\to\ (A,Z+2)\to 2\ e^-,$$ 
have been done. This process is possible only
if neutrinos are massive and Majorana particles. The matrix element of the process 
is proportional to the effective Majorana mass 
$\langle m\rangle=\sum \eta_i U_{ei}^2 m_i$. 
Uncertainties in the precise value of upper limits are important  since 
they  depend on  theoretical calculations of   nuclear matrix elements.  
From the non-observation of 
$(\beta\beta)_{0\nu}$ the Heidelberg-Moscow Ge experiment \cite{heidelberg}
 draws the limit 
$\mid\langle m\rangle\mid <0.5-1.5 $ eV (90\% CL). 
In the next years it is expected an increase in  sensitivity  allowing 
limits down  the
$\mid\langle m\rangle\mid \sim 0.1$ eV level.

Many short-baseline (SBL) 
neutrino oscillation experiments with reactor and accelerator neutrinos did not find any evidence of neutrino oscillations. 
For example experiments looking for 
$\overline{\nu}_e\to\overline{\nu}_e$ or 
${\nu}_\mu\to{\nu}_\mu$ 
dissaperance \cite{Bugey,CCFR} or oscillations 
$\overline{\nu}_\mu\to\overline{\nu}_e$ \cite{E776,CCFR}.

\begin{figure}[ht]
\centering
\mbox{\psfig{file=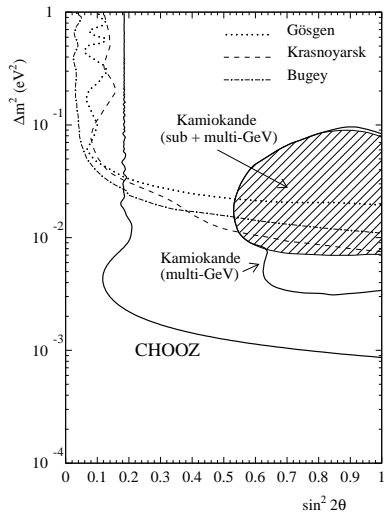,height=7cm}}
\vspace{0cm}
\caption{The 90\% C.L. exclusion plot for CHOOZ, 
compared with previous experimental 
limits and with the KAMIOKANDE allowed region.} 
\label{fCHOOZ}
\end{figure}

The first reactor long-baseline neutrino oscillation experiment 
CHOOZ found  no evidence for neutrino oscillations in the $\nuebar$
disappearance mode \cite{CHOOZ}.
Their results  imply an exclusion region  
in the plane of the two-generation mixing parameters 
(with normal or sterile  neutrinos) 
given
approximately by $\Delta m^2 > 0.9~10^{-3}\units{eV^2}$ for maximum mixing and 
$\sinsq > 0.18$ for large $\dmsq$, as shown in Fig.(\ref{fCHOOZ}).
CHOOZ results are important for the atmospheric deficit problem: as it is 
seen in Fig.(\ref{fCHOOZ}) they are incompatible with 
an  $\nue\to \numu$ oscillation hypothesis for the solution of the 
atmospheric problem.

Los Alamos LSND experiment  has reported  
indications of  possible
$\overline{\nu}_\mu\to\overline{\nu}_e$ oscillations \cite{LSND}. 
They search for 
$\overline{\nu}_e$'s in excess  of  the number expected from conventional 
sources at a liquid scintillator detector located 30 m from a proton
  beam dump at LAMPF. 
A $\overline{\nu_e}$ signal has been detected via the reaction
$\overline{\nu}_e p\to e^+ n$ with $e^+$ energy between 36 and 60 MeV, 
followed by a $\gamma$ from $n p\to d\gamma$ (2.2 MeV). 
A total $\overline{\nu}_e$ excess of $51.8^{+18.7}_{-16.9}\pm 8.0$ 
events has been obtained.
If this excess is
 attributed to neutrino oscillations of the type 
$\overline{\nu}_\mu\to\overline{\nu}_e$, 
it corresponds 
to an oscillation probability of 
$3.1\pm 1.0\pm 0.5\times 10^{-3}$. 
The results of a similar search for 
$\nu_\mu\to \nu_e$ 
oscillations where the $\nu_e$ are detected via the CC reaction 
$C(\nu_e,e^-) X$ 
provide a value for the corresponding oscillation probability of 
$2.6\pm 1.0\pm 0.5\times 10^{-3}$.

\begin{figure}[ht]
\centering
%
\psfig{file=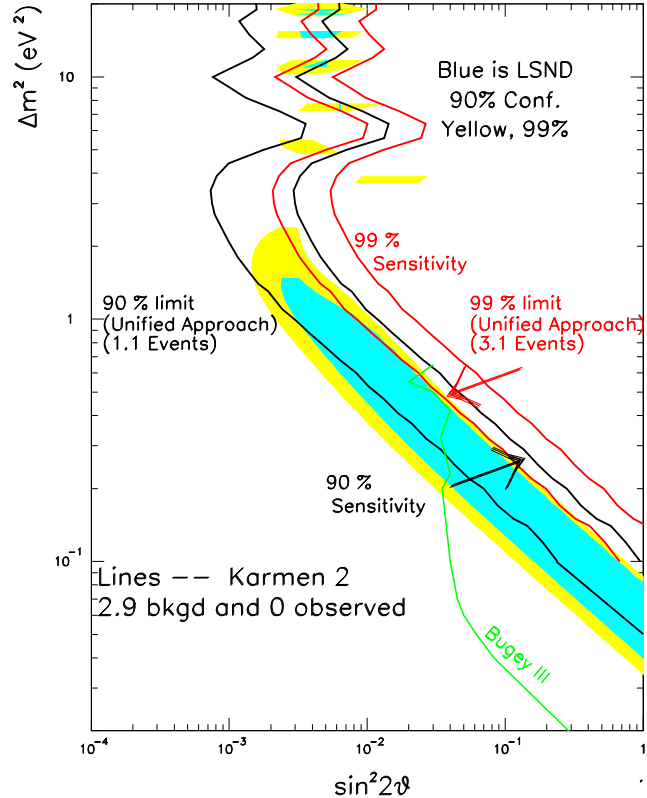,height=13cm}
\vspace{-2cm}
\caption{ The LSND 1993-97 likelihood regions along 
with KARMEN2 limits (unified approach)
together with a priori sensitivity.}
\label{fLSND}
\end{figure}

The LSND result has not been confirmed by the KARMEN experiment. 
The KARMEN experiment (Rutherford Laboratories), following a similar 
experimental setup as  LSND,
searches for $\bar\nu_e$ produced by
$\bar\nu_\mu\to\bar\nu_e$ oscillations at a mean distance of 17.6 m. 
The time structure of the neutrino beam
is important for the identification of the neutrino induced reactions
and for the suppression of the cosmic ray background.
Systematic time anomalies not completely understood has been reported.
The 1990-1995 and the latest 1997-1998 KARMEN
   data showed inconclusive results. They found  no events,
with an expected background of $ 2.88 \pm 0.13 $ events,
 for  
$\overline{\nu}_\mu\to\overline{\nu}_e$ oscillations, however the 
positive LSND result in this channel could not be excluded in total 
either \cite{KARMEN}. At the end of  1999, the 
KARMEN sensitivity is expected to be able to 
exclude the whole parameter region of evidence suggested by LSND if 
no oscillation signal were  found (Fig.\ref{fLSND}).
The first phase of a third pion beam dump 
 experiment designed to set the LSND-KARMEN controversy 
has been approved to run at Fermilab. 
Phase I of ''BooNe'' ( MiniBooNe) expects a 10 $\sigma$ signal 
($\sim 1000 $ events)
 and thus will make a decisive statement either proving or 
ruling it out. Plans are to run early 2001. Additionally, there is a letter
 of intent of a similar experiment to be carried out at the CERN PS 
\cite{BOON,CERNPS}.

\subsection{Solar neutrinos}

Indications in the favor of neutrino oscillations were found in ''all'' 
solar neutrino experiments:
The Homestake Cl radiochemical experiment with sensitivity down 
 to the lower energy parts of the $^{8}$B neutrino spectrum and to the 
higher $^{7}$ Be line.
The two radiochemical $^{71}$Ga experiments, SAGE and GALLEX, which are 
 sensitive to the low energy pp neutrinos and above and the 
water Cerenkov experiments Kamiokande and Super-Kamiokande (SK) which 
 can observe only the highest energy $^{8}$ B neutrinos. Water 
 Cerenkov experiments in addition demonstrate directly that the 
neutrinos come from the Sun showing that recoil 
electrons are scattered in the direction along 
the sun-earth axis.

Two important points to remark are:  
a) The prediction of the existence of a global neutrino deficit is 
hard to modify due to the constraint of the solar luminosity on pp 
neutrinos detected at SAGE-GALLEX.  
b) The different  experiments are sensitive to neutrinos with different energy ranges and 
 combined yield spectroscopic information on the neutrino flux.
Intermediate energy neutrinos arise from intermediate steps 
of the thermonuclear solar cycle. It may not be 
 impossible to reduce the flux from the last step ($^{8}$B), for example 
by  reducing temperature of the center of the Sun, but it seems extremely 
 hard to reduce neutrinos from $^7$Be to a large extent, while 
keeping a reduction of $^8$B neutrinos production to a modest amount.
If minimal standard electroweak theory is correct, the shape of the 
$^8$ B neutrino energy spectrum is independent of all solar influences
 to very high accuracy.

\begin{figure}[ht]
\centering
\psfig{file=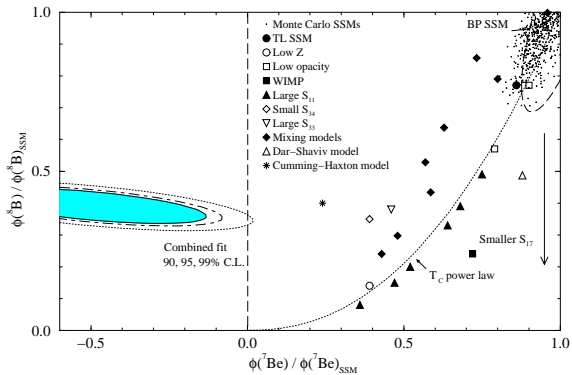,height=10cm}
\vspace{-2cm}
\caption{
The severity of the problem with astrophysical solutions.
The constraints on $^8$B and $^7$Be fluxes (considered as 
free parameters) from the combined 
Cl, Ga, and \v{C}erenkov experiments
( 90, 95, and 99\% C.L.)   are shown.
The best fit solutions are obtained for unphysical values.
Diverse
standard and nonstandard solar models
are shown.
[From Hata and Langacker, Ref.(\protect\cite{hata} and 
references therein.]}
\label{SSM1}
\end{figure}

\begin{figure}[ht]
\centering
\psfig{file=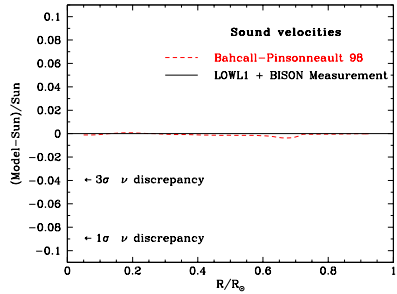,height=9cm}
\vspace{-3cm}
\caption{ 
The excellent agreement
between the calculated (solar model BP98) 
and the measured (Sun) sound speeds. The fractional error is much smaller
than generic fractional changes in the model, 0.03 to 0.08, that might significantly affect
the solar neutrino predictions. 
[Adapted from Christensen-Dalsgaard, Ref.\protect\cite{dalsgard}, as 
it appears in \protect\cite{berezinsky}.] }
\label{SSM2}
\end{figure}

Unless the experiments are seriously in error, there must be 
 some problems with either our understanding of the 
Sun  or neutrinos. Clearly, the SSM cannot account for the data 
(see Fig.\ref{SSM1})
and possible 
highly nonstandard solar models are strongly constrained by 
heliosysmology studies (See Fig.(\ref{SSM2})).

There are at least two reasonable particle physics explanations 
that could account for the suppression of intermediate energy neutrinos. The 
first one, neutrino oscillations in vacuum, 
requires a large mixing angle and a seemingly unnatural 
fine tuning of neutrino oscillation length with the Sun-Earth distance 
 for intermediate energy neutrinos. 
The second possibility, level-crossing effect  
oscillations in presence of solar matter and/or 
magnetic fields of regular and/or chaotic nature (MSW, RSFP), requires 
 no fine tuning either for mixing parameter or neutrino mass difference 
 to cause a selective large reduction of the neutrino flux. This mechanism
 explains naturally the suppression of intermediate energy neutrinos, leaving 
 the low energy pp neutrino flux intact and high 
energy $^8$B neutrinos only loosely suppressed. 
Concrete range of parameters obtained including  the latest SK data will 
be showed in the next section.

\subsection{The SK detector and Results.}

The  high precision and high statistics 
 Super-Kamiokande (SK) experiment initiated operation in 
 April 1996.
A few words about the detector itself.
SK is a 50-kiloton water Cerenkov detector 
located near  the old Kamiokande detector under a 
mean overburden of  2700 meter-water-equivalent.
The effective fiducial volume is  $22.5$ kt.
It is a well understood, well calibrated detector. 
The accuracy of the absolute energy scale is estimated to be 
$\pm 2.4\%$ based on several independent calibration sources:
cosmic ray through-going and stopping muons, muon decay 
electrons, the invariant mass of $\pi^0$'s produced by neutrino interactions, radioactive source calibration, 
and, as a novelty in neutrino experiments, 
a 5-16 MeV electron LINAC. 
In addition to the ability of recording higher 
statistics in less time, 
due to the much larger dimensions of the detector, SK can contain 
multi-GeV muon events  making possible for the first 
time a measurement of the spectrum of $\mu$-like events up to 
$\sim 8-10 $ GeV/c.

The results from the first 
504 days of data from SK (results presented recently at the 
Neutrino98 conference \cite{nu98}), combined with data from 
earlier experiments provide important constraints on the MSW and 
vacuum oscillation solutions for the solar neutrino problem (SNP).
In the next paragraphs we will present a summary of the results 
presented in that conference together with initial analysis in 
the framework of neutrino oscillations.

The most robust  results of the solar neutrino experiments so far are the 
total observed rates.
The most recent data on rates are summarized in table~(\ref{t1}).
Total rates alone indicate
 that the $\nu_e$ energy spectrum from the Sun is distorted.
The SSM flux predictions are inconsistent with the observed 
rates in solar 
neutrino experiments at approximately 
the 20$\sigma$ level. 
Furtherly,
 there is no linear combination of neutrino fluxes 
that can fit the available data at the 3$\sigma$ level 
(Fig.(\ref{SSM1})).

\begin{table}[h]
\centering
\small
\begin{tabular}{|c|c|c|c|}
 \hline\vspace{0.1cm}
Experiment      & Target & E. Th. (MeV)  & $S_{Data}/S_{SSM}\ (1\sigma)$
\\[0.1cm] \hline
Homestake  & $^{37}$ Cl & 0.8 &  $0.33\pm 0.029  $              \\[0.1cm]
Kamiokande & H$_2$O  & $\sim 7.5$ &   $0.54\pm 0.07$  \\[0.1cm]
SAGE       & $^71$Ga & 0.2 &                 $    0.52\pm 0.06 $    \\[0.1cm]
GALLEX     & $^71$Ga &0.2 &    $  0.60\pm 0.06 $  \\[0.1cm]
SK (504 days)& H$_2$O & $\sim$ 6.5  &$ 0.474\pm 0.020$
\\ \hline       
\end{tabular}
\caption{Neutrino event rates measured by solar neutrino experiments,
        and corresponding predictions from the  SSM
        (see Ref.\protect\cite{bah2}  and references therein, we take the INT normalization for the 
   SSM data ( $1\sigma$ errors).}
\label{t1}
\end{table}

From a two-flavor  analysis of the total event rates in the 
ClAr, SAGE,GALLEX and SK experiments the best $\chi^2$ fit considering
 active neutrino oscillations is obtained for 
$\Delta m^2=5.4\times 10^{-6}$ eV$^2, \sin^2 2\theta=6.0\times 10^{-3}$
(the so called small mixing angle solution, SMA).
Other local $\chi^2$ minima exist. The large mixing angle solution
(LMA) occurs at 
$\Delta m^2=1.8\times 10^{-5}$ eV$^2, \sin^2 2\theta=0.76$,
the LOW solution (lower probability, low mass), at
$\Delta m^2=7.9\times 10^{-8}$ eV$^2, \sin^2 2\theta=0.96$.
The vacuum oscillation solution occurs at 
$\Delta m^2=8.0\times 10^{-11}$ eV$^2, \sin^2 2\theta=0.75$. 
At this extremely low value for the  mass difference the MSW effect is 
inopperant.

For oscillations involving sterile neutrinos 
(the  matter effective potential is modified in this case) 
the LMA and LOW solutions are 
not allowed and only the  (only slightly modified) 
SMA solution together with the vacuum solution  are still possible.

More sophisticated analysis including more than two neutrino 
species are not available but they would not change so much the 
previous picture while introducing a much larger 
technical difficulty. Analysis which  consider  neutrino oscillations in  
presence of magnetic fields, the RSFP effect, 
have also been presented . Typically, they yield   solutions with 
$\dms\sim 10^{-7}-10^{-8}$ eV$^2$ for both small and large mixing angles.
RSFP solutions are much more ambiguous than pure MSW solutions because of 
neccesity to introduce additional  free parameters in order to model 
 the largely unknown intensity and profile of solar magnetic fields. 
The recognition of the random nature of solar convective  
  fields 
and recent theoretical developments in the treatment of Schroedinger 
random equations have 
partially aliviated this situation, allowing the obtention of 
SNP solutions without the neccesity of a detailed model description
 (see recent analysis in \cite{tor2,bykov,tor5}).  
In addition, random RSFP models predict the production of 
a sizeable quantity of electron antineutrinos.  
Presently, antineutrino searches with negative  
results in Kamiokande and SK   are welcome because restrict significantly
the, uncomfortably large, parameter space of RSFP models. In the future such antineutrinos
 could be identified both in SK or in SNO setting the Majorana nature 
of the neutrino (\cite{tor2,bykov,tor5}).

\begin{figure}[ht]
\centering
\psfig{file=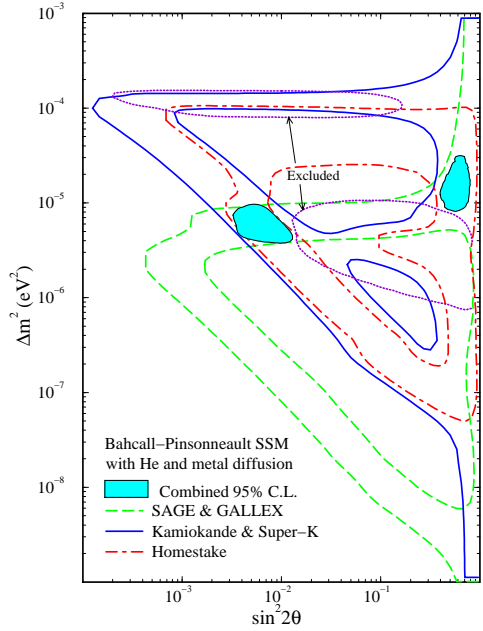,height=11cm}
\vspace{-1cm}
\caption{
The result of the MSW parameter space (shaded regions) allowed by the
combined observations at 95\% C.L.\ aassuming the Bahcall-Pinsonneault
SSM with He diffusion.  The constraints from Homestake, combined
Kamiokande and Super-Kamiokande, and combined SAGE and GALLEX are
shown by the dot-dashed, solid, and dashed lines, respectively.  Also
shown are the regions excluded by the Kamiokande spectrum and
day-night data (dotted lines).
[From Hata and Langacker, Ref.(\protect\cite{hata} and 
references therein.]}
\label{SOLAR1}
\end{figure}

If MSW oscillations are effective, for a certain range of  neutrino 
parameters the observed event rate will depend upon the 
zenith angle of the Sun (Earth matter regeneration effect). 
After 504 days of data still, due to lack of statistics, the most 
 robust  estimator of zenith angle dependence by now is the Day-night
 (or up-down) asymmetry, A. The experimental estimation is:
\begin{eqnarray}
A\equiv\frac{D-N}{D+N}=-0.023\pm0.020\pm 0.014, \quad (E_{recoil}>6.5\ {\rm MeV}).
\end{eqnarray}
The difference is small and not statistically significant 
but it is in the direction that would be expected from 
regeneration at Earth (the Sun is apparently neutrino brighter at night).
Taken alone the small value observed for A excludes a large part of 
the  parameter region  that is allowed if only the 
total rates would be considered (see Fig.(\ref{SOLAR1})).

\begin{figure}[ht] 
\centering
\psfig{file=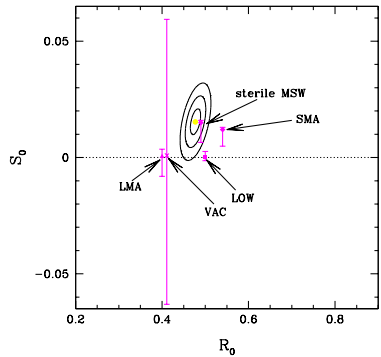,height=6cm}
\vspace{-0.5cm}
\caption[]{Deviation from an undistorted energy spectrum.
The $1,2,3\sigma$ allowed regions are shown  in the figure.
The ratio of the observed counting rate as a function of electron 
recoil energy~\protect\cite{nu98} to the expected
undistorted energy spectrum was fit to a linear function of
energy, with intercept $R_0$ and slope $S_0$. 
The five oscillation 
solutions  
SMA active and sterile, LMA, LOW,
and vacuum oscillations,
 all provide
acceptable fits to the data, although the fits are not particularly
good.
[From Bahcall and Krastev, Ref.(\protect\cite{bah2})].}
\label{SOLAR2}
\end{figure}

From the independence from astrophysical causes, the shape of the 
 neutrino spectrum determines the shape of the recoil electron 
 energy spectrum produced by neutrino-electron scattering in the detector.
All the 
neutrino oscillation solutions (SMA,LMA,LOW and Vacuum) provide 
acceptable, although indeed not excellent fits to the recoil 
energy spectrum. The simplest test is to investigate whether the 
ratio, R, of the observed to the standard energy spectrum is a
 constant. The fit of the ratio R to a linear function yields 
 slope values which are incompatible at 99\% CL with the 
hypothesis of no distortion
(see Figs.(\ref{SOLAR1}-\ref{SOLAR2})).

In the case 
 where all data, 
the total rates, the zenith-angle 
dependence and the recoil energy spectrum, 
is combined the best-fit solution is 
almost identical 
to what is obtained for the rates only case. 
For other solutions, only the SMA and  vacuum 
solution survives (at the 99\% CL). 
The LMA and the LOW solutions are, albeit marginally, ruled out
\cite{bah2}.

\begin{figure}[ht]
\centering
\psfig{file=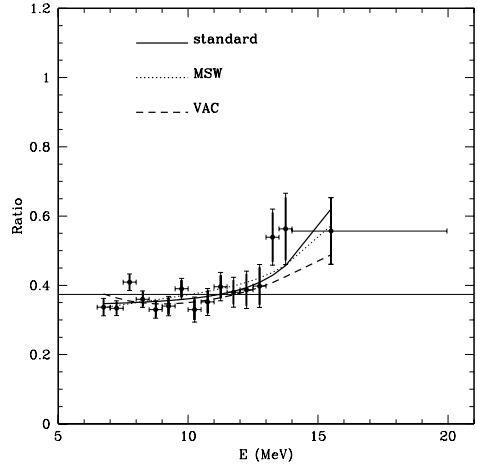,height=7cm}
\vspace{0cm}
\caption{
Nuclear physics calculations of the rate of the hep reaction are uncertain. 
The  figure  show the results for the predicted energy spectrum that is 
measured by SK (\protect\cite{nu98}). 
The total flux of hep neutrinos was varied to
obtain the best-fit for each scenario. 
The calculated curves are
global fits to all of the data, the chlorine, GALLEX, SAGE, 
and SK total
event rates, the SK energy spectrum and Day-Night asymmetry. 
[Figure reproduced from Ref.(\protect\cite{bah5})].}
\label{SOLAR6}
\end{figure}

A small but significant discrepancy appears
when comparing  the predictions from the 
global best fits for the  energy spectrum 
at high energies
($E_{\nu}\gsim 13 $ MeV)
with the SK results presented in 
Neutrino98~\cite{nu98}. 
From this discrepancy it has been speculated that
hep neutrinos may affect the solar neutrino energy spectrum.
Presently
low energy nuclear physics calculations of the rate of the hep reaction are highly uncertain (a factor of six is allowed). 
Coincidence between expected and measured ratios is improved 
when the hep flux is  allowed to vary 
as a free parameter (see Fig.(\ref{SOLAR6})).

\begin{figure}[ht]
\begin{center}
\centering
\epsfig{file=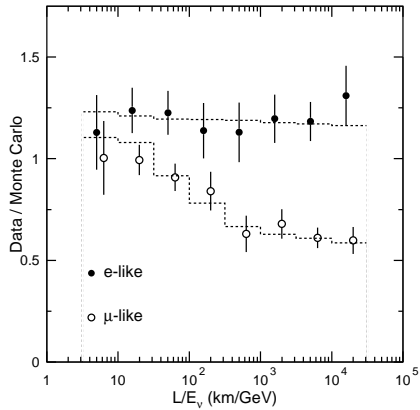,height=6cm} 
\vspace{0cm}
\caption{The SK multi-GeV data sample.
The ratio of the number of FC (fully contained) data events to FC Monte Carlo events versus
reconstructed $L/E_\nu$. Points: absence of oscillations. 
Dashed lines: 
expected shape for $\numu \leftrightarrow \nutau$ at
$\dms=2.2\times10^{-3} $eV$^2$ and $\sstt=1$.
The slight $L/E_\nu$ dependence for $e$-like events is
due to contamination (2-7\%) of $\nu_\mu$ CC interactions.
[From Ref.\protect\cite{sk2}]
}
\label{ATM4}
\end{center}
\end{figure}

\clearpage

\subsection{Atmospheric neutrinos}

Atmospheric neutrinos are the decay products of hadronic 
showers produced
 by cosmic ray interactions in the atmosphere. 
The experimental ratio R
$$R\equiv \left ( \mu/e\right )_{DATA}/\left ( \mu/e\right )_{MC} $$
where $\mu/e$ denotes the ratio of the numbers of $\mu$-like to 
$e$-like neutrino interactions observed in the data or predicted by the simulation
has been measured as an estimator of the atmospheric neutrino 
flavor ratio 
$(\nu_\mu+\overline{\nu}_\mu)/(\nu_e+\overline{\nu}_e).$ 
The individual  absolute neutrino flux calculation has a $20\%$ uncertainty.
The flux ratio  has been calculated however 
 to an accuracy of better then 
$5\%$ in the range $0.1-10$ GeV. 
The calculated flux ratio has a value 
of about 2 for energies $<$ 1 GeV and increases with increasing neutrino energy.
For neutrino energies 
higher than a few GeV, the fluxes of upward and 
downward going neutrinos are expected to be nearly equal; 
geomagnetic field
 effects at these energies are expected to be small because of the relative
large geomagnetic 
rigidity of the primary cosmic rays that produce these 
neutrinos.

Anomalous, statistically significant, 
low values of the ratio R  has been obtained previously 
in the water Cerenkov detectors Kamiokande and IMB-3 
and the  Soudan-2 for ``sub-GeV'' events (E$_{vis}< 1$ GeV). 
The NUSEX and Frejus experiments have  reported results 
consistent with no deviation from unity with  smaller data samples.
Kamiokande  experiment observed a value of R smaller than unity 
in the  multi-GeV (E$_{vis}>$1 GeV) energy region  as well as a 
dependence of this ratio on the zenith angle. 
IMB-3, with a smaller data sample,
 reported inconclusive results in a similar energy 
 range but  they are 
 not in contradiction with Kamiokande observations.

\begin{figure}[ht]
\centerline{\protect\hbox{\psfig{file=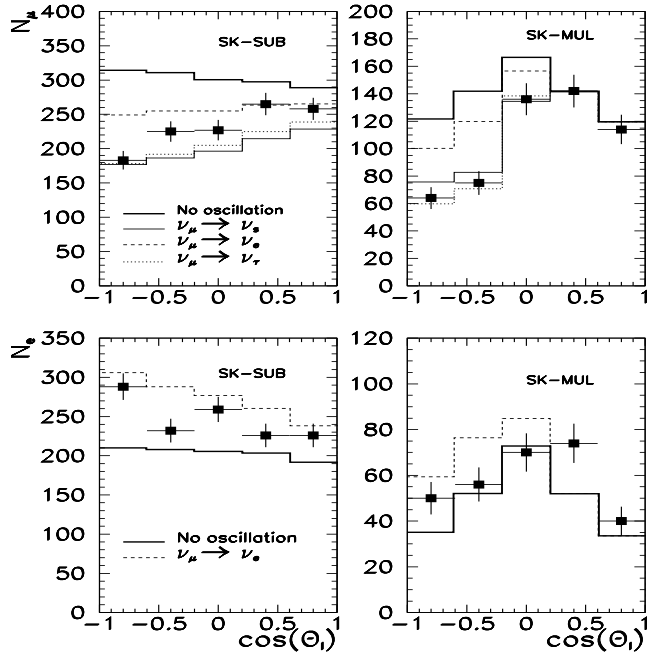,width=0.6\textwidth,height=0.4\textheight}}}
\caption{Angular distribution for Super-Kamiokande electron-like and muon-
like sub-GeV and multi-GeV events. Predictions in the
absence of oscillation (thick solid line), $\nu_\mu \to
\nu_s$ (thin solid line), $\nu_\mu \to \nu_e$ (dashed line) and
$\nu_\mu \to \nu_\tau$ (dotted line).  The errors displayed in the
experimental points is only statistical.
[From  Ref.\protect\cite{atm1}]}
\label{ATM8}  
\end{figure}

SuperKamiokande (SK) recent results are completely consistent
with previous results but they are more accurate than before.
Specially significant improvements in accuracy have been obtained
in measuring the zenith angular dependence of the neutrino events.
They see that the flux of muon neutrinos going up is smaller than
that of downgoing neutrinos:
In the sub-GeV range ($E_{vis}< 1.33$ GeV), from an exposure of $22.5$ Kiloton-years of the SK detector the measured ratio R is:
$$R=0.61\pm0.03\pm 0.05. $$
It is not possible 
 to determine from data, whether the observed deviation of R is due to an electron excess of a muon deficit. 
The distribution of R with momentum in the 
sub-GeV range is consistent with a flat distribution 
within the statistical error as it happens 
with zenith angle distributions (see right plots in Fig.(\ref{ATM8})). 
In the multi-GeV range, it has been obtained (for a similar 
exposure) a ratio R which is 
slightly higher than at lower energies $R=0.66\pm0.06\pm 0.08. $
For e-like events, the data is apparently consistent with 
MC. For $\mu$-like events there is a clear discrepancy between
 measurement and simulation.

A strong distortion in the shape of the $\mu$-like event 
 zenith angle distribution was observed 
(Plots (\ref{ATM4}-\ref{ATM8})). 
The angular correlation between the neutrino direction 
and the produced 
charged lepton direction is much better at higher energies ( $\sim 15^0-20^0$): 
the zenith angle distribution of leptons reflects rather accurately that of the neutrinos in this case.
The ratio of the number of
upward to downward $\mu$-like events was found to be 
$$(N_{up}/N_{down})^\mu_{Data}=0.52^{+0.07}_{-0.06}\pm 0.01$$
while the expected value is practically one:
$(N_{up}/N_{down})^\mu_{MC}=0.98\pm 0.03\pm 0.2.$
The validity of the results  has been  tested by 
measuring the azimuth angle distribution of the 
incoming neutrinos, which is insensitive to 
a possible influence from neutrino oscillations. 
This shape agreed with MC predictions which were nearly flat.

\begin{figure}[ht]
\begin{center}
\centering
 \epsfig{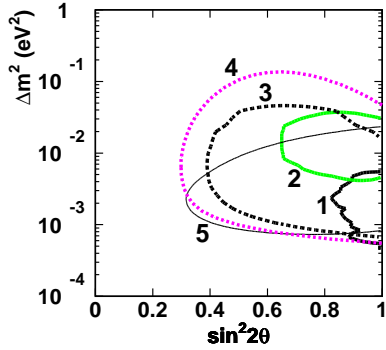}
\end{center}
\caption{ The allowed neutrino oscillation parameter regions 
obtained by Kamiokande and SK ( 90\% C.L..).
 (1) and (2):
the regions obtained by contained event analyses from Super-Kamiokande
and Kamiokande, respectively. 
(3) and (4):  upward through-going muons from
SK and Kamiokande, respectively. 
(5) stopping/trough-going ratio analysis of 
upward going muons from SK.
[From  Ref.\protect\cite{atm2}]}
\label{ATM6}
\end{figure}

The most obvious solution to the observed discrepancy is 
 $\nu_\mu\to\nu_\tau$ flavor neutrino oscillations.  
This fits well to the  angular distribution, 
since  there is a large 
difference in the neutrino path-length between upward-going 
($\sim 10^{4}$ Km) and downward-going ($\sim 20$ Km), a zenith 
angle dependence of R can be interpreted as  evidence  for 
neutrino oscillations. 
Oscillation into sterile neutrinos, $\nu_\mu\to \nu_s$, 
is also a good explanation consistent with data. 
$\nu_\mu-\nu_e$ oscillations does not fit however so well, they would also 
conflict laboratory measurements (CHOOZ,Figs.(\ref{fCHOOZ}-\ref{ATM6})). 
Apart from
neutrino oscillation, no other consistent explanation has been proposed.

Evidence for oscillations equals 
evidence for   non-zero neutrino mass within the standard neutrino 
theory.
The allowed neutrino oscillation parameter regions 
obtained by Kamiokande and SK  from different analysis are 
shown in Fig.(\ref{ATM6}). 
The best fit is obtained for squared mass differences in 
the range 
$10^{-2} - 10^{-3}$ eV$^2$, and very large
mixing. Unless
there is no fine tuning, this suggests  a 
neutrino mass of the order of 0.1 eV. 
Such a mass implies the neutrino energy density in 
the universe to be 0.001 of the critical density 
which is too small to have cosmological consequences. 
This is of course a very rough argument:
specific models, however, may allow larger neutrino masses
quite naturally.

\begin{figure}[ht]
\centering
\psfig{file=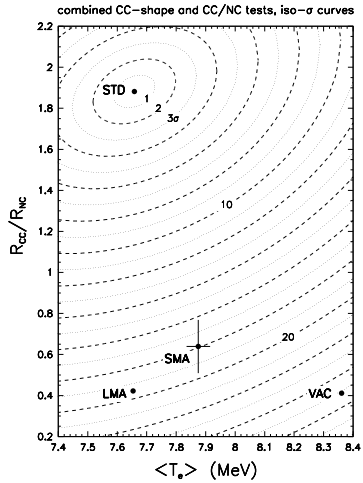,height=8cm}
\vspace{0cm}
\caption{Conclusive probes of lepton number violation in solar neutrino 
experiments. 
Iso-sigma contours for the SNO for the
combined CC-shape and CC/NC test, for the representative oscillation cases. 
Iso-sigma contours 
 for the combined CC-shape and CC/NC test, for
representative oscillation cases. 
STD = standard (no oscillation); SMA =
small mixing angle (MSW); LMA = large mixing angle (MSW); VAC = vacuum
oscillation. 
[From  Ref.\protect\cite{bahcall2}]}
\label{FUT3}
\end{figure}

\clearpage

\subsection{Global multi-fold analysis and the neccesity for 
sterile neutrinos.}

From the individual  analysis  of the data available from neutrino 
experiments, it follows that there exist three different 
scales of neutrino mass squared differences and two different 
ranges of small and maximal mixing angles, namely:
\begin{eqnarray}
\Delta m_{sun}^2&\sim  10^{-5}-10^{-8} 
\ eV^2\ ,& \sin^2 2\theta\sim 7\times 10^{-3}
(MSW,RSFP), \\
&\sim 10^{-10} \ eV^2,& \sin^2 2\theta\sim 0.8-0.9\ (Vac.); \\ 
\Delta m_{Atm}^2&\sim  5\times 10^{-3} \ eV^2,&\  \sin^2 2\theta\sim 1\\
\Delta m_{LSND}^2&\sim  3\times 10^{-1}-2\ eV^2&\  \sin^2 2\theta\sim 10^{-3}-10^{-2}.
\label{e1001}
\end{eqnarray}
Fortunely for the sake of simplicity the neutrino mass scale relevant 
 for HDM is roughly similar to the LSND one. The  introduction of the 
former  would not change any further conclusion.
But for the same reason, the definitive refutation of LSND results by 
KARMEN or future experiments does not help completely in simplifying the 
task  of finding a consistent framework for all the 
neutrino phenomenology.

Any combination of experimental data which involves only of the 
two mass scales can be fitted within a three family scenario, but 
solving simultaneously the solar and atmospheric problems requires generally 
some unwelcome fine tuning of parameters at the $10^{-2}$ level.
The detailed analysis of Ref.\cite{BILENKY} 
 obtains for example that
 solutions with 3 neutrino families which are compatible with the 
results from SBL inclusive experiments, LSND and solar neutrino experiments are possible.
The problem arises when 
 one add the results from CHOOZ, which rule out large 
atmospheric $\nu_\mu \nu_e$ transitions and zenith dependence from 
SK atmospheric data one comes to the neccesity of 
consideration of schemes with four massive neutrinos
  including  a light sterile neutrino.
Among the numerous possibilities,
 complete mass hierarchy of four neutrinos is not favored by existing 
data \cite{BILENKY} nor four-neutrino mass spectra with one 
neutrino mass separated from the group of the three close masses by the 
''LSND gap'' ($\sim$ 1 ) eV. One is left with two possible 
 options where  two double-folded groups of close masses 
are separated by a $\sim 1 $ eV gap:
\begin{eqnarray}
&(A)&\ \underbrace{\overbrace{\nu_e\to\nu_s: \ m_1< m_2}^{sun}<< \overbrace{\nu_\mu\to \nu_\tau:\ m_3< m_4}^{atm}}_{LSND\sim 1 eV} \\[0.1cm]
&(B)&\ \underbrace{\overbrace{\nu_e\to \nu_\tau: \ m_1< m_2}^{sun}<< \overbrace{\nu_\mu\to\nu_s:\ m_3< m_4}^{atm}}_{LSND\sim 1 eV}.  
\end{eqnarray}

The two models would be distinguishable from the detailed analysis of future 
 solar and atmospheric experiments. 
For example they may be tested 
combining future precise  recoil electron spectrum in $\nu e\to \nu e$ 
measured in SK and SNO \cite{SNO} with
the  SNO spectrum measured in CC absorption.
The SNO experiment 
(a 1000 t heavy water under-mine  detector) 
will measure the rates of the charged (CC) and neutral (NC) current reactions
induced by solar neutrinos in deuterium:
\begin{eqnarray}
\nu_e + d \rightarrow p+p+e^-\quad({\rm CC\ absorption}),&&
\nu_x + d \rightarrow p+n+\nu_x\quad({\rm NC\ dissociation}).
\label{reactionNC}
\end{eqnarray}
including the determination of the electron recoil energy in the CC 
reaction. Only the more energetic $^8$B solar neutrinos
are expected to be detected since the expected
SNO threshold  for CC events is an electron kinetic energy
of about 5 MeV and the physical threshold for NC dissociation is the 
binding energy of the deuteron, $E_b= 2.225$ MeV.
If the (B) model it is true one 
expects $\phi^{CC}/\phi^{NC}\sim 0.5$ while in the 
(A) model the ratio would be $\sim 1$.
The schemes (A) and (B) give different predictions for the neutrino mass 
 measured in tritium $\beta$-decay and for the effective Majorana mass
 observed in neutrinoless double $\beta$ decay. Respectively we have 
$\mid \langle m\rangle\mid < m_4$ (A) or $<< m_4$ (B).
Thus, if scheme (A) is realized in nature this kind of experiments can see the 
effect of the LSND neutrino mass.

From the classical LEP requirement $N_\nu^{act}=2.994\pm 0.012$ \cite{PDG98}, it is clear that 
the fourth neutrino  should be  a $SU(2)\otimes U(1)$ singlet in order 
to ensure that does not affect the invisible Z decay width.
The presence of additional weakly interacting light particles, such as a 
light sterile $\nu_s$, is constrained by BBN since it would enter into 
equilibrium with the active neutrinos via neutrino oscillations.
The limit 
$\Delta m^2 \sin^2 2\theta< 3\times 10^{-6}$ eV$^2$ should be 
fulfilled in principle. 
However systematical uncertainties in the 
derivation of the BBN bound make any bound too unreliable to be 
taken at face value and can eventually be avoided \cite{foot}. 
Taking the most restrictive options 
(giving $N_\nu^{eff}< 3.5$) only the  (A) scheme is allowed, one where the 
sterile neutrino is mainly mixed with the electron neutrino.
In the lest restrictive case ($N_\nu^{eff}< 4.5$) 
both type of models would be allowed.

\section{Conclusions and future perspectives.}

The theoretical challenges that  the present phenomenological situation 
 offers are two at least: to understand origin and, 
very particularly,the lightness of the sterile 
neutrino (apparently requiring a radiatively generated mass) and 
to account for the maximal neutrino mixing indicated by the 
atmospheric data which is at odd from which one could expect from 
considerations of the mixing in the quark sector.
Actually, the existence of light sterile neutrinos could even be 
beneficial in diverse astrophysical and cosmological scenarios (
 supernova nucleosynthesis, hot dark matter, lepton and 
baryon asymmetries for example).

In the last years different indications in favor of nonzero neutrino masses 
and mixing angles have been found.
These evidences include 
 four solar experiments clearly 
demonstrating an anomaly compared  to the predictions of the 
Standard Solar Model (SSM) and a number of other atmospheric 
experiments, including a high statistics, well calibrated one, demonstrating
 a quite different anomaly at the Earth scale.

One could argue that if we are already  beyond the 
stage of having only
''circumstantial evidence for new physics'',
 we are still however a long way from having 
''conclusive proof of new physics''. 
Evidence for new physics does not mean the same as evidence for neutrino 
oscillations but there exists  a significant 
 case for neutrino oscillations and hence neutrino masses and mixing 
 as ''one'', indeed the most serious candidate, explanation of the data.
One of the possible alternatives is that one or more of the 
experiments will turn out to be wrong. This is possible and even probable,
 but it is little probable that with all the evidence accumulated by now 
 all the experiments turn out to be simultaneously wrong.

Many neutrino experiments are taking data, are going to start or are 
under preparation: solar neutrino experiments
(SNO and Borexino are of major interest, also HERON, HELLAZ, ICARUS, GNO and  others);
LBL reactor (CHOOZ, Palo Verde, KamLand) and accelerator experiments
(K2K, MINOS, ICARUS and others);
SBL experiments  (LSND, KARMEN, BooNE and many others). 
The important problem for any  next generation experiment is to find
specific and unambiguous 
experimental probe that the ''anomalies'' which has been found 
 are indeed signals of neutrino oscillations and to distinguish among the 
different neutrino oscillation possibilities (this is specially important 
in the Solar case).
Among these probes, we could include:
\begin{itemize}
\item Perhaps the  most direct test of 
SM deviation:   to measure the 
ratio of the flux of $\nu_e$'s (via a  CC interaction) to the 
flux of neutrinos of all types ($\nu_e + \nu_\mu + \nu_\tau$,
determined 
by  NC interactions). This measurement will be 
done hopefully by the SNO experiment in the near future.
See Fig.(\ref{FUT3}).

\item Statistically significant demonstration of an
 energy-dependent modification of the shape of the electron neutrino 
spectrum arriving at Earth. Besides observing distortion in the shape of
$^8$B neutrinos, it will be very important to make direct 
measurements of the $^7$Be (Borexino experiment) 
and pp (HERON,HELLAZ) neutrinos.
\item Improved observation of a zenith angle effect
in atmospheric experiments or their equivalent,  a
 day-night effect  in solar experiments.
\item And least, but by no means the least, independent confirmation by 
one or more accelerator experiments.
\end{itemize}
There is a high probability that in the near future  we should know much more 
than now about the fundamental properties of neutrinos and their 
 masses, mixing and their own nature whether Dirac or Majorana.

\vspace{1cm}
{\bf Acknowledgments}

The author has been supported by DGICYT under Grant 
 PB95-1077 and by  a DGICYT-MEC contract  at Univ. de Valencia.

\end{document}